\documentclass[12pt,aps,prb,preprint]{revtex4}   % style for Physical Review B and AJP ar %similar

\usepackage{amsmath}    % need for subequations
\usepackage{graphicx}   % for figures

\begin{document}

\title{Unpacking Symbolic Equations in Introductory Physics}
\author{Eugene T. Torigoe}
 \affiliation{Thiel College, Greenville, PA 16125}
 \email{etorigoe@thiel.edu}   %optional
\date{\today}

\begin{abstract}
Symbolic equations are one of the many representations used in physics.  Understanding these representations is important for students because they are how students access knowledge in physics.  In this paper I build off of the work by Redish and Kuo~\cite{Redish-15}, which described the cultural differences between how math is used in math class and physics class, and the work by Fredlund et al~\cite{Fredlund-14}, which described the importance of unpacking representations for physics student.  I will describe how differences in the goals of numeric and symbolic problem solving lead to different set of affordances.  In particular, the inability to distinguish variables, knowns, and unknowns in symbolic problem solving is a benefit when describing a generalized physical system.  I also present evidence that observed errors when trying to solve symbolic problems are due to students acting on inappropriate cues associated with numeric problem solving.  
\end{abstract}

\maketitle
\section{Introduction}

There has been a great deal of research about the difficulties students encounter when using math in introductory physics.  Many physics instructors find that their students are unable to apply the appropriate mathematical tools even when they have passed the required prerequisite math courses.  One tool researchers have used to study this difficulty is the math diagnostic exam.  The purpose of these exams is to measure the mathematical skills students possess when they begin introductory physics.  The results allow the instructor to tailor his or her lessons to the weaknesses of the students.  The results of these exams have been shown to correlate to student success in introductory physics.~\cite{Hudson, Halloun, Meltzer}

These mathematical difficulties are mirrored in the math education research literature.  Researchers have studied difficulties students encounter when they make the transition from arithmetic to algebra.~\cite{Kieran-90, Kieran-92, Filloy, Goodson-Espy}  While arithmetic focuses mainly on numeric computation, algebra subsumes arithmetic and also incorporates symbolic representation.  While this research mostly focuses on elementary, and high school students taking algebra courses, this work has also been expanded, and found similar difficulties with college students.~\cite{Trigueros-03, Clement-82, Cohen-05}  There are many college age students who have difficulties applying basic algebraic concepts.  

My prior work has focused on measuring the difficulties students encounter while trying to solve symbolic physics questions in the context of an introductory physics class.  Studies of introductory physics students found that they performed far better on questions with numbers, than otherwise analogous questions using only symbols.~\cite{Torigoe-06, Torigoe-11}  These difficulties mirror the types of difficulties seen by students first learning algebra, and making a transition from arithmetic to algebra. These results bolster the claim by some physics instructors that many students are not sufficiently prepared mathematically for the tasks required in introductory physics.

While the evidence of difficulty is compelling, Redish and Kuo~\cite{Redish-15} argue that the inability of students to transfer from math class to physics class is not the appropriate focus.  They argue that while the math in math class may be superficially similar to the math used in physics class, it is actually quite different.  These two systems originate from culturally different disciplines with different goals.

\begin{quote}\emph{
More precisely, in the culture of physics, the use of mathematical expressions is complex, because the ancillary physical meaning of symbols is used to convey information omitted from the mathematical structure of the equation. This is because we have a different purpose for the math: to model real physical systems.~\cite{Redish-15}}
\end{quote}

In physics the underlying goal of the mathematical equations is to represent some aspect of physical reality.  They argue that the transfer of mathematical skills learned in math classes into physics class is of limited value as a target for instructors or researchers.

\begin{quote}\emph{
Even if students have learned the relevant mathematical tools in their math courses, they still need to learn a component of physics expertise not present in math class tying those formal mathematical tools to physical meaning.~\cite{Redish-15}}
\end{quote}

Because math in physics is used as a representational tool reflecting the properties of a physical system, the way mathematical symbols are interpreted is completely different in physics classes than in math classes.

Expert representational systems are designed by experts for the community of experts that use them.  Such complex representational systems can be opaque to novices.  Fredlund et al~\cite{Fredlund-14} call the process where representations become generalized, and obscured, the rationalization process. 
 
\begin{quote}\emph{
The rationalization process has led to a more generalized representation. However, from a student point of view, using such generalized representations is even more problematic since it calls for an in-depth understanding of how these representations relate to the particular situations at hand~\cite{Fredlund-14}}
\end{quote}

The process of rationalization has two main mechanisms: nominalization and rank shifting.  Fredlund borrows both concepts from the linguistics literature.  Nominalization acts to transform verbs into nouns, which increases the flexibility of the language used.  For example, the transformation of the statement ``kinetic energy is conserved" into the ``conservation of kinetic energy", leads to a more general and flexible concept of conservation.  Rank shifting transforms a more complex unit of language into a less complex one.  The simplified notation often makes the meaning less accessible to novices because interpreting the meaning of these rank shifted representation depends on expert defined implicit cues.

Learning physics occurs within the context of the representations defined by physics experts.  According to Fredlund et al, in order for students to gain access to this disciplinary knowledge, they must first be able to appropriately use and interpret these various representations, which they call the appreciation of the disciplinary affordances of the representation.

\begin{quote}\emph{
I define the disciplinary affordances of a given representations as the inherent potential of that representation to provide access to disciplinary knowledge.  Thus, it is these disciplinary affordances that enable certain representations to become legitimate within a discipline such a physics.  Physics learning then, involves coming to appreciate the disciplinary affordances of representations.~\cite{Fredlund-12}}
\end{quote}

Fredlund et al argue that a primary goal of the instructor is the unpacking of these expert systems to novice students.  This is a difficult process for experts within the discipline because the function of the representations have been internalized, and automated.

\begin{quote}\emph{
In many cases teachers have become so familiar with the disciplinary representations that they use that they no longer “notice” the learning hurdles involved in interpreting the intended meaning of those representations.~\cite{Fredlund-14}}
\end{quote}

The rationalization process can be demonstrated by the representations used in math classes.  Sfard describes historically common processes, similar to nominalization, in which mathematical processes evolve into mathematical structures/objects.~\cite{Sfard}  For example, the square root of a negative number was first conceived of as a type of process, but later conceived as an object, namely an imaginary number.  As a nominalized object, the concept has much more flexibility and utility.  Now it can be acted upon by different processes or combined to form other more complex quantities.  

When students make the transition from arithmetic to algebra the equations undergo a process similar to rank shifting.  In arithmetic all equations are computational equations, which cue the computation of a particular quantity within the equation.  In algebra the equations may also contain structural objects that can be used in the same ways as numbers would.  This is a complex process because the structural interpretation does not replace the process interpretation but instead becomes a dual interpretation that can switch back and forth depending on the context.  Gray and Tall~\cite{Gray-94, Tall-01} have coined the term procept to describe this phenomena.  

\begin{quote}\emph{
An elementary procept is the amalgam of three components: a process that produces a mathematical object, and a symbol that represents either the process or the object.~\cite{Gray-94}}
\end{quote}

The procept is a type of rank shifting because instead of having a complex notation that clearly demarcates the process from the conceptual object, the notation is left ambiguous as a way for mathematicians to flexibly move from one interpretation to the other without the burden of separate notations.

\begin{quote}\emph{
Instead of having to cope consciously with the duality of concept and process, the good mathematician thinks ambiguously about the symbolism for product and process. I contend that the mathematician simplifies matters by replacing the cognitive complexity of process-concept duality by the notational convenience of process-product ambiguity.~\cite{Gray-94}}
\end{quote}

In this paper I examine the disciplinary affordances of symbolic physics equations, and I contrast that with the disciplinary affordances of equations using numbers.  I argue that problems involving numbers form a representational system with different goals than the representational system for problems involving only symbols.  While a mapping can be created between numeric and symbolic solutions, the notational systems serve different purposes and therefore have developed different representational affordances.

I also adopt the theoretical framework of resources.~\cite{Hammer}  Under this framework, the context of the activity and the cues that one perceives activate a web of resources that one brings to bear on a particular activity.  These resources can includes pieces of factual knowledge as well as epistemological resources about the nature of the task.  From this perspective a particular error may be due to the absence of a resource or the failure to activate the appropriate resources.  The activation of a resource can act as a cue to activate other resources.  And so for an expert, the features of a particular problem very likely activate a whole web of resources appropriate to the problem.

I will argue that one aspect of difficulty with symbolic equations is the surface similarity to numeric problems.  Further, that when novice students see symbolic equations, they activate a web of resources appropriate for solving numeric problems, which are inappropriate, and fail to sensitize them to the cues for the resources appropriate for symbolic problem solving.

\section{The goals of numeric and symbolic problem solving}

The evolution of expert representational systems is driven by the affordance of particular representations to serve the goals of the discipline.  In this section I will discuss the differences in the goals between numeric and symbolic problem solving, and speculate on how the affordances of the representations have evolved to suit these goals.  

The use of numbers in physics is a reflection of the importance of experimental evidence within the discipline.  This is the connection between quantitative measurements with mathematical models.  An equation can be used with particular numeric measurements with units to make a computation that can be directly compared to another measurement of the world.  This affords the testing of mathematical models to the world, as well as predictions of future measurements based on a mathematical model.

The use of purely symbolic equations in physics serves a different but complementary purpose in the discipline.  Symbolic equations demonstrate the underlying relationships between quantities within the mathematical model.  They allow the expert to have a deeper understanding of the mathematical model, and its connection to an abstract and generalized physical system.  Symbolic equations are not constrained to a particular experimental context, but are representative of a generalized physical system, which has embedded within it a range of possible physical systems.  While numeric computations obscure the relations among various variables, symbolic equations display these relationships.  Even in cases when a particular variable is canceled out of the equation it is an indicator that it has no relationship to the other independent variables in that context.

One consequence of this goal is that symbolic physics equations must allow flexibility in how the symbols in an equation are to be interpreted.  The equations must function as a model of a specific physical context, as when it is used in problem solving, but they must also afford a description of the class of similar contexts, a generalized system, in which the symbols are variable.  For example, when experts examine limiting cases to compare their conceptual expectations of the physical system, and the mathematical model when it has been stretched to its limits.

The differences in these goals lead to a different affordances for the two mathematical representations.  In addition, I will argue that experts in the discipline have different sets of resources and resource activation cues when working on different types of problems.

\section{Unpacking equations used during numeric problem solving}

Equations with numbers are embedded in a context of a specific physical situation.  The goal is to incorporate one or more numeric measurements into a mathematical model to predict the numeric value of a target quantity, which can be compared to an actual measurement.

With this goal in mind the affordances of the representation must clearly distinguish the structure of the mathematical model, the unknown quantities, and the known quantities.  The known quantities could be numeric measurements or physical constants.  The structure of the mathematical model is represented as a purely symbolic equation. For example,

\begin{quote}
	$v_{f} = v_{o} + a t$
\end{quote}

This particular model for motion applies to systems traveling at a constant acceleration.   Numeric measurements and constant with units are plugged into the equation as follows:

\begin{quote}
  $20 m/s = 5 m/s + a (3 s)$
\end{quote}

In this equation the numbers in the equation clearly represent the known quantities, and the target quantity stands out as the remaining non-numeric symbol (excluding units).  The target quantity is isolated, and the known quantities can be combined to give a single numeric result.

\begin{quote}
	$a=3 m/s^{2}$ 
\end{quote}

If this is not the final target quantity, then it is at least clear that it can serve as an intermediate target, in which the quantity can be treated as a known quantity to be plugged into other equations.

This system of letters and numbers serves the purpose of calculating a value for a quantity from a mathematical model in the context of a particular physical situation.  It clearly demarcates the mathematical model from the specific known and unknown quantities.

Numeric problem solving is associated with a sensitivity to particular cues.  These cues trigger the activation of a web of resources to perform the task at hand.  In particular, in many cases numeric problems involve an equation that contains a single unknown symbol from the mathematical model that hasn’t been replaced by a number.  A single unknown is to be isolated in the equation to determine its value.

\section{Unpacking equations used during symbolic problem solving}

The goal of symbolic equations is to represent the relationships between variables, and the properties of a generalized physical system.  In the context of problem solving, this goal leads to a conflict between a representation of the specific physical system under examination, and a generalized form of that system.  A problem is situated within a specific physical system, with known quantities that stand in for specific measurements, and unknown quantities which serve as target quantities.  But to serve as a mathematical model of a generalized version of the system, the same symbols must also represent variables.  While the lack of a notation to distinguish a known, unknown, or variable quantity obscures the meaning for novices, that notational ambiguity is an affordance of the representation for the expert who prefers flexibility.  This is essentially the same argument by Gray and Tall for the procept idea.  For example, in the equation $v_{f} = v_{o} + a t$ there is no way of distinguishing it as a general form of a mathematical model, a mixture of known and unknown quantities relating to the specific physical system in the problem, or a generalized version of the system.

When compared to the notational system for numeric problem solving there are clear differences.  First, it is not clear which of the symbols should serve as the target unknown.  The determination of which symbol should be isolated is dependent on the context of the problem, and perhaps the knowledge of which quantities are easily experimentally measured, and which are not.  Second, there is no clear demarcation between the general mathematical model and the quantities relating to the physical system.

Further complicating the interpretation of the equation, a single equation can contain multiple symbols that are associated with different objects, intervals of space, or intervals of time.  This complication occurs when equations associated with different object or intervals are combined into a single equation.  This is a type of confusion in symbolic problem solving that is not present in numeric problem solving, because only numbers are passed from one equation to the next.  When solving symbolic problems one must be aware of the associations of each of the symbols being used.

This level of information about the relation of various variables with one another allows the expert to make frequent check-ins with the generalized physical system.  The expert can engage in various methods of interrogating the mathematical model with their expectations of the generalized physical system.

\section{Inappropriate resource activation: how students view symbolic equations through the lens of numeric problem solving.}

There is a growing body of evidence that students are able to solve numeric problems, but not the analogous symbolic version of that problem.~\cite{Torigoe-06, Torigoe-11, Kortemeyer, Brahmia}.  Clearly the format of the question does not influence the students' knowledge of physics.  I argue that the source of the discrepancy is due to the activation of an inappropriate set of resources for solving symbolic problems.

I have argued in the preceding sections that the goals, cues, and resources appropriate for numeric problem solving are inappropriate for symbolic problem solving.  So errors are bound to occur when a student applies the procedures appropriate for numeric problem solving in an attempt to solve a symbolic problem.  Both numeric and symbolic problem solving begin with a symbolic equation that is representative of the mathematical model to be applied.  In both numeric and symbolic problems the equations are the same.  The next step involves the identification of known quantities, which can include numeric measurements or numeric constants.  This is difficult to do in the symbolic equation because the notation does not distinguish known, unknown, and variable quantities.  The following rules work for numeric problem solving but do not work for symbolic problem solving.

\begin{itemize}
	\item Symbolic equations are a general mathematical model
	\item The remaining non-numeric symbol is the target unknown to be isolated
\end{itemize}

It would make sense, then to see errors related to the following:

\begin{itemize}
\item Treating all symbolic equations as generalized mathematical models
\item Treating (non-numeric) symbols representing known quantities as unknown quantities 
\end{itemize}

The cues for resource activation that students rely on to solve numeric problems fail them.  After plugging in the numeric values, they cannot rely on the normal visual cue of the remaining non-numeric symbol to identify the unknown.  They may not be able to distinguish the general mathematical model from an equation relating specifically to the problem at hand.  When solving for a target unknown, the expression will not be easily identified as a known quantity without first identifying its constituent parts as also known quantities.

\section{Evidence of confusion with symbolic problems}

Speak-aloud problem solving interviews with thirteen introductory physics students were performed.  During the interviews each student was first given the symbolic version of a question to solve while speaking aloud about their method to reach the solution.  Whether correct or incorrect, the subjects were asked questions to gauge their understanding of the symbols in the problem.  If a subject had difficulties with the symbolic version, they were asked to solve the numeric version of the same question. If that subject was then able to solve the numeric version, then the subject was then asked to use their numeric solution to find the correct symbolic expression. The students were never told whether they found the correct or incorrect result.  

\begin{figure}
\begin{center}
\scalebox{0.5}{\includegraphics{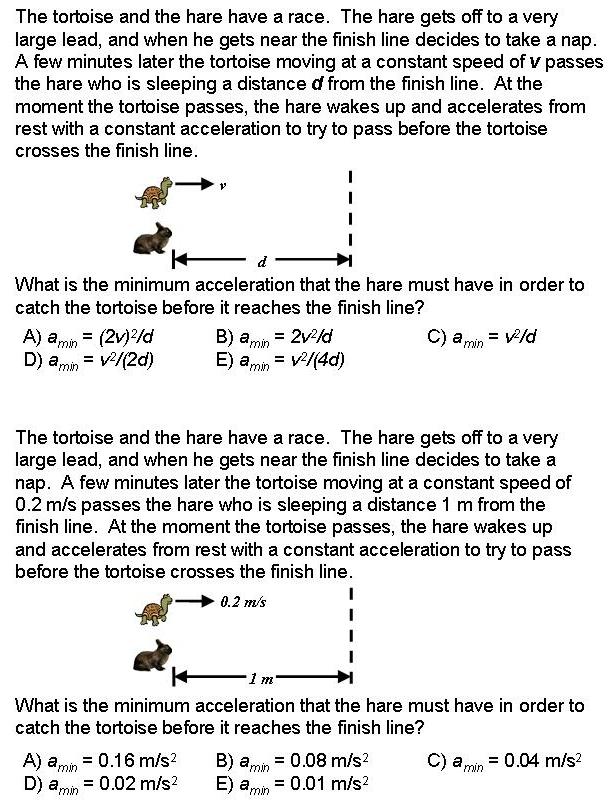}}
\caption{\label{fig:TandH}The numeric and symbolic versions of the
\emph{Tortoise and Hare} question.  This question is analogous to the \emph{Bank
robber} question from the 2006 study.~\cite{Torigoe-06}}
\end{center}
\end{figure}

\begin{figure}
\begin{center}
\scalebox{0.5}{\includegraphics{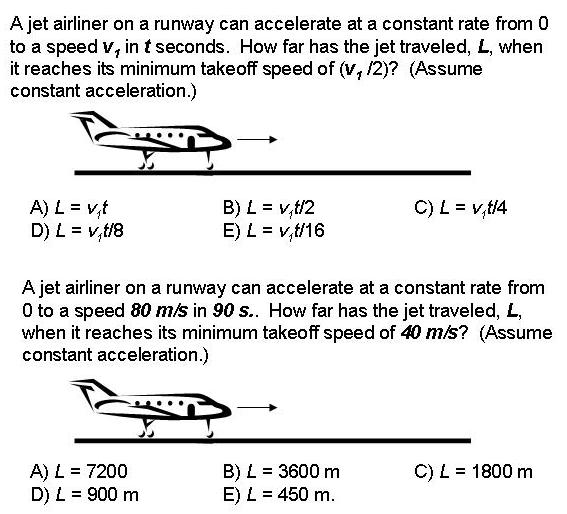}}
\caption{\label{fig:Plane}The numeric and symbolic versions of the
\emph{Airliner} question.  This question is analogous to the \emph{A car can go}
question from the 2006 study.~\cite{Torigoe-06}}
\end{center}
\end{figure}

The physics questions used in this study were modified versions of the questions used in the earlier final exam study.~\cite{Torigoe-06} The questions used during the interviews can be seen in Figures \ref{fig:TandH} and \ref{fig:Plane}.  The structure of the questions were the same but with different surface features.  

During these interviews it was common to observe a students who were unable to correctly solve the symbolic problem, but when given the numeric problem immediately afterward, could find their earlier mistakes and could easily find the correct answer.

\subsection{Confusion of known and unknown symbols}

Many students exhibited difficulties distinguishing known and unknown symbols when working on the symbolic versions.  During the interviews students often lost track of the known and unknown quantities.  For example, while solving the Tortoise and Hare symbolic problem three students started with the equations $d = v_0t$ and $d = \tfrac{1}{2}a t^2$, and incorrectly eliminated the known quantity ($d$) leaving the two unknowns ($a$ and $t$). As a result they ended up with one equation and two unknowns.  All three were surprised when their final result was not one of the answer options.  The following quote was from one of these students.

\begin{quotation}\emph{
\textbf{Subject:} [Starts with equations $d = v_0t$ and $d = \tfrac{1}{2}a t^2$] um now solving for $a$, I actually solved for, so $v_ot = \tfrac{1}{2}a t^2$, let's start over, $vt$, $2vt = at^2$ divided $t^2$ [on one side of the equation], divided by $t^2$ [on the other side of the equation], equals $a$, cross those guys out, $2v_o/t = a$, and that would give you none of the answers given, which stinks!... [Answer contains the unknowns $a$ and $t$ because he eliminated the known quantity $d$]}
\end{quotation}

These types of errors were not observed in the numeric versions because known and unknown quantities were easily identified by either using a number or a letter, respectively.

\subsection{Variable confusion}
In order to apply a general mathematical model to a specific system, quantities representing the system must be plugged into the general equation.  While this is easy with numbers, it is much more difficult with purely symbolic equations.  There is evidence from the interviews some students never specified the general mathematical model to match the properties of the system described in the problem.  Those students seemed to interpret any purely symbolic equation as a general mathematical model.

This can be seen from student errors in the Tortoise and the Hare problem (Figure~\ref{fig:TandH}).  In this problem a Tortoise moving with a constant speed passes a Hare at rest a certain distance from the finish line.  The instant the Tortoise passes, the Hare accelerates toward the finish line.  Given the speed of the Tortoise and the distance from the finish line the question asks you to find the minimum acceleration needed for the Hare to catch up to the Tortoise before the finish line.

The most common error was to use the equation $v_f^2 = v_i^2 + 2a\Delta x$, to get the incorrect result $a = v^2/(2L)$. This is incorrect because the symbol $v$ is given as the velocity of the Tortoise, but is used as if it were the velocity of the Hare when it reaches the finish line.  There were five students who were observed to make this error.  Those five students were questioned in an attempt to determine if they had a consistent interpretation of the symbol. Two of the five said that the symbol represented the velocity of the Tortoise, and showed great flexibility in their interpretations of that symbol.  In the following exchange, one of those students switched their interpretation of v in less than a minute.

\begin{quote}\emph{
[Student selected the answer $a_{min} = v^2/(2L)$]\\
\textbf{Interviewer:} OK and what does $v$ represent? [points to the selected answer]\\
\textbf{Subject:} umm, the veloc, the \textbf{constant velocity of the tortoise}\\
\textbf{Interviewer:} Alright, so let's say that instead of this question asking for the minimum acceleration, asked you to find ... um... the vel, the final velocity of the hare, do you think you could write down an equation for the final velocity of the hare when it reaches the finish line?\\
\textbf{Subject:}  umm, I think I would just, probably rearrange this equation [referring to her final answer of $a_{min} = v^2/(2L)$]\\
\textbf{Interviewer:} OK\\
\textbf{Subject:} Because it, blah, the acceleration does not change, I mean its constantly accelerating, but in this scenario it’s still the minimum acceleration, and distance doesn't change, so I would  just rearrange the equation to find the final velocity.\\
\textbf{Interviewer:} And that would be the \textbf{final velocity of the hare?}\\
\textbf{Subject:} \textbf{Right}}\end{quote}

These two students did not seem to hold any firm connection between the symbol $v$ and any particular property of the physical system.  They treated the equation in an overly general way.

There is some evidence from the Airliner question (shown in Figure~\ref{fig:Plane}) that the plugging in numbers acted as a cue to specify the variable and apply the symbol with particular associations.  In this problem an airliner on a runway accelerates at a constant rate from rest.  The airliner's final velocity and the time to reach that velocity are given, and the question asks you to find the distance traveled when the airliner reaches half the final velocity.

One of the observed errors for this problem was when the students used the symbol $t$, which is defined as the time to reach the final velocity, as the time to reach half the final velocity instead.  Two interview subjects made this specific error.  These two students started by using the general equation $x = x_i + v_i t + \tfrac{1}{2}at^2$. They made the replacements $x_{i} = 0$, $v_{i} = 0$, $x = L$, and $a = v/t$ to get $L = \tfrac{1}{2}(v/t)t^2$.  The students simplified this to the equation $L = vt/2$. Unfortunately the symbol $t$ is used to represent two different times.  The $t$ from the general equation should properly represent the time to reach a speed $v/2$, and the $t$ in the acceleration equation should represent the time the reach a velocity $v$. The students erred when they canceled the t's after combining the two equations.

Even though they made this error on the symbolic version, both were able to correctly solve the numeric version. The act of plugging in numbers seemed to act as a cue to specify the meaning of the variable they were replacing.  Both attempted the same procedure as they had in the symbolic version, but as they were about to plug in a value for the time into the kinematics equation, they realized that it was necessary to solve for the specific time when the jet airliner reached a speed of 40 m/s (half the final speed), and that it was inappropriate to cancel the $t$'s as they had in the symbolic version.

\begin{quote}\emph{
\textbf{Subject:} Umm, $x= \tfrac{1}{2}a t^2$, and then $a = v/t$ ... so when I plugged in the $x$ equals, uhhh, $a = v/t$ in my equation $x = \tfrac{1}{2}a t^2$, \textbf{I crossed out the times, but [the t in the acceleration equation] was for when it was 90 [s] and [the t in the general equation] is when, we don't know how long it took}.  So maybe I should...figure out... how long it takes for the plane to get to 40 m/s...[Subject then correctly solves the numeric version]\\
\textbf{Interviewer:} OK so how confident do you feel about that?\\
\textbf{Subject:} Umm, I was pretty confident, but I kind of got sidestepped over what time I should use, so I went to the side and solved for it.}\end{quote}

Another aspect of the numeric problem solving that the students seemed to benefit from was the isolation of symbols with different associations.  In this example the inclusion of numbers allowed the students to isolate each meaning of the symbol $t$ from the other definition by the use of separate numeric equations.

\section{Discussion}
I use Fredlund et al's rationalization framework to make the claim that numeric and symbolic notational systems, while similar, have different affordances within the discipline of physics.  The goal of the mathematical representations used during numeric problem solving is to relate a mathematical model with a specific physical system.  Either in the sense of using the mathematical model to make predictions about the physical system, or using the physical system to test the applicability of the mathematical model.  The goal of the symbolic mathematical representation is to identify the relationship between variables, as well as the relationship of the equation to a generalized version of the physical system being analyzed.  As a result of the different goals they have developed different affordances, resources, and cues for resource activation.

It is very common for experts in physics to prefer to solve problems symbolically, even when numbers are given in the problem.  One reason this might be the case is because the symbolic equations give more information about the mathematical model of the generalized system.  Symbolic equations allow for a broader analysis of the physical situation, and if required, the model of the generalized system can then be specified with the numeric measurements and constants of the specific system being studied.

I also use this distinction to explain observed difficulties that students have while attempting to solve symbolic problems.  My claim is that the difficulty students have with symbolic problem solving originates from the application of cues and resources for numeric problem solving to symbolic problems, where such a set of resources leads to confusion and error.  

An understanding of the affordances of these representations should not be expected to be understood by students on the first day they take physics.  The use of math in physics is culturally specific to the discipline of physics.  Fredlund emphasizes that to have access to disciplinary knowledge requires that students understand the affordances of the representations used within the discipline.  One of the goals of teaching physics is to unpack the representations which have been rationalized over time.

One aspect of this unpacking is to help students appreciate the benefits of each representation’s affordances.  To do this a match must exist between the affordances of the representation and the goals of the instructional activity.  Numeric problem solving should be used to either make predictions on a physical system, or to test the applicability of a mathematical model using measurements of the physical system.  And symbolic problems should be used to find the properties of the generalized physical system, including the relationship between variables, and/or the testing of the model using limiting cases.  In activities where there is a mismatch between the affordances of the representation being used and the result of the activity, then it seems inevitable that students will be confused about the utility of the representations.  

Physics instructors commonly advise introductory physics students to solve all physics problems symbolically, and then plug in the given values as the last step.  If the goal of the activity is the determination of a numeric value, then the students will be justifiably confused by the necessity of having solved the problem symbolically.  The students will not see the benefit of such a method unless they fully utilize the affordances of the representation.  In the case of symbolic problem solving the students must be able to use the symbolic expression to gain an understanding of the generalized physical system.  Prompts to connect the symbolic equations to generalized physical system may serve to show the benefits of working symbolically.

Traditional text book problems that use the correct numeric result as the metric for success advantage numeric problem solving over symbolic problem solving.  Based on the evidence that students have much more difficulty with symbolic versions of questions than the analogous numeric versions, students will likely to experience success when working with numbers, and failure and frustration when working with symbols.  Even if the instructor tells them to work symbolically before plugging in numbers, many students will find that they are much more successful plugging in numbers as soon as they can.  This is a major barrier to students learning the benefits of symbolic problem solving.

One way of approaching the issue is the change the metric of success in problem solving so that the advantages of symbolic problem solving over numeric problem solving are clear.  Some possibilities include:

\begin{itemize}
	\item Examination of the generalized system using multiple sets of possible measured values
	\item Examine Limiting Cases
	\item Interpretation of the relationship between variables
	\item Interpreting the association between the symbolic equation and the physical system
\end{itemize}

The chain of association between the physical system and the mathematical equation is an affordance of symbolic equations in physics that is important to emphasize to students.  Redish and Kuo suggest that instructors focus on the connection of the equations to the physical system, which differentiates the use of math in physics from its use in math class.  Specifically they suggest that instruction should begin with physical intuition, and then connect physical intuition and the mathematical models.  

\begin{quotation}\emph{
It might well be preferable to ‘‘teach physics standing on your head’’ by beginning with the physical meaning and creating a chain of association to the math, both strengthening the students’ skills of ‘‘seeing physical meaning’’ in equations and helping them develop the epistemological stance that equations in physics should be interpreted physically.~\cite{Redish-15}}
\end{quotation}

Brahmia et al.~\cite{Brahmia-2}, describe the blending and translating between mathematical equations and the physical world as ``mathematization".  To promote this type of thinking they employ invention tasks where the students are encouraged to create their own mathematical descriptions of carefully chosen physical phenomena.   

In order for students to appreciate symbolic problem solving, they must understand the affordances of the representation.  The representations need to be unpacked so that the students can understand how the physical system is represented by the mathematical model.  While the state of a symbol as a variable, known, or unknown may appear to be a hindrance to the use of symbolic equations, it affords the flexibility in the equation to describe a more generalized version of the physical system being studied.  It allows for more correspondences between the physical system and the mathematical model, than numeric equations.

\section{Conclusion}
I make the claim that numeric and symbolic problem solving serve different goals within the discipline of physics.  These disparate goals lead to different representational affordances.  As instructors it is important that we unpack these representations, and demonstrate the value of each representation.  Of particular importance is the flexibility of symbolic equations to represent a generalized physical system.  Such a representation allows one to more easily connect physical intuition to the mathematical model.

\end{document}